\documentclass[10pt,a4paper]{article}
\usepackage{amsmath,latexsym,amssymb,amsfonts}
\usepackage[dvips]{graphicx}
\usepackage{bm}

\begin{document}
\title{\textbf{Euclidean quantum gravity and\\ stochastic inflation}}
\author{\textsc{Dong-il Hwang}$^{a,b}$
,\;\; \textsc{Bum-Hoon Lee}$^{a}$
,\;\; \textsc{Ewan D. Stewart}$^{b}$
,\\ \textsc{Dong-han Yeom}$^{a}$
\;\; and\;\; \textsc{Heeseung Zoe}$^{b,c}$\\
\textit{$^{a}$\small{Center for Quantum Spacetime, Sogang University, Seoul 121-742, Republic of Korea}}\\
\textit{$^{b}$\small{Department of Physics, KAIST, Daejeon 305-701, Republic of Korea}}\\
\textit{$^{c}$\small{Division of General Studies, UNIST, Ulsan 689-798, Republic of Korea}}
}
\maketitle

\begin{abstract}
In this paper, we compare dispersions of a scalar field in Euclidean quantum gravity with stochastic inflation. We use Einstein gravity and a minimally coupled scalar field with a quadratic potential. We restrict our attention to small mass and small field cases.

In the Euclidean approach, we introduce the ground state wave function which is approximated by instantons. We used a numerical technique to find instantons that satisfy classicality. In the stochastic approach, we introduce the probability distribution of Hubble patches that can be approximated by locally homogeneous universes down to a smoothing scale. We assume that the ground state wave function should correspond to the stationary state of the probability distribution of the stochastic universe.

By comparing the dispersion of both approaches, we conclude three main results. (1) For a statistical distribution with a certain value, we can find a corresponding instanton in the Euclidean side, and it should be a complex-valued instanton. (2) The size of the universe of the Euclidean approach corresponds to the smoothing scale of the stochastic side; the universe is homogeneous up to the Euclidean instanton. (3) In addition, as the mass increases up to a critical value, both approaches break at the same time. Hence, generation of classical inhomogeneity in the stochastic approach and the instability of classicality in the Euclidean approach are related.
\end{abstract}

\newpage

\tableofcontents

\newpage

\section{Introduction}

In the semi-classical approach of gravity, we quantize fields in a classical metric background.
Although the semi-classical approach cannot be the fundamental theory, it could give a useful guideline on how to approach to the problems of quantum gravity.
One of the examples is Hawking radiation in black hole physics \cite{Hawking:1974sw}.
To study the nature of Hawking radiation, there are two useful approaches.
One way is to treat quantum fluctuations of a field on the classical curved background \cite{Birrell:1982ix}.
The other way is to use the Euclidean path integral and obtain some thermodynamic quantities by the Wick rotation \cite{Gibbons:1994cg}.

In the same spirit, we can tackle cosmological problems with two different approaches.
As an application of the former (quantum field theory in a curved background),
we can illustrate \textit{stochastic inflation} \cite{Vilenkin:1983}\cite{Linde:1993xx}\cite{Tolley:2008na}.
Here, we can divide the Fourier modes into those shorter than the Hubble radius and longer than the Hubble radius.
As long as the potential is broad and the field value is small, we can approximate that the longer wavelengths behave like a locally homogeneous and classical field while the shorter wavelengths act as a Gaussian random noise to the longer wavelengths. Therefore, the locally homogeneous scalar field will behave like a Brownian particle.
This randomly walking field can be described by the \textit{Langevin equation}.
In the entire (inhomogeneous) universe, there are many Hubble patches that behave like this,
and one may define the probability distribution of the fields for each of the Hubble patches.
The master equation of this probability distribution is the \textit{Fokker-Planck equation}.

As an application of the latter (Euclidean quantum gravity), it is useful to study the path integral. The ground state solution of the Schrodinger equation with gravity, so-called the \textit{Wheeler-DeWitt equation}, is described by the Euclidean path integral \cite{Hartle:1983ai}. It is not easy to calculate the whole path integral; however, if we only restrict to the on-shell solutions (instantons), then we can approximately obtain the wave function. In many contexts, instantons are useful to study non-perturbative phenomena of the universe. For example, the Coleman-DeLuccia instanton \cite{Coleman:1980aw} is useful to describe the inhomogeneous vacuum decaying process, while the Hawking-Moss instanton \cite{Hawking:1981fz} is used to describe the homogeneous tunneling process.

One interesting observation is that the stationary solution of the Fokker-Planck equation corresponds to the Hawking-Moss instanton \cite{Linde:1993xx}. The question is that \textit{is this correspondence an accident or not?} If there is no gravity, we can find many examples that a solution of the Fokker-Planck equation is indeed a solution of the Schrodinger equation \cite{LeB}. However, it is not trivial if we include gravitation. Can we still extend this relation in the presence of gravity?

In this paper, keeping in mind this problem, we suggest three questions:
\begin{itemize}
\item If the stationary solution of the Fokker-Planck equation and the ground state of the superposition of the instantons are (approximately) the same, what is the Euclidean instanton that corresponds to a certain state for the stationary solution?
\item We already know that if the field is almost static, two approaches give the same results. Then, what will happen if the field begins to move? Even for this limit, are these two approaches the same?
\item If we increase the curvature of the potential (increase the mass around the local minimum), the stochastic description will breakdown. What is the corresponding phenomena in the Euclidean side?
\end{itemize}
To investigate these issues, in Section~\ref{sec:sto} and Section~\ref{sec:euc}, we discuss the basics of the stochastic approach and the Euclidean approach. In Section~\ref{sec:the}, we compare two approaches and discuss the first observations of both approaches. In Section~\ref{sec:com}, we compare the details of two approaches, and we will answer the previous questions. Finally, in Section~\ref{sec:con}, we summarize our discussions.

\section{\label{sec:sto}Stochastic approach}

Let us assume the Einstein gravity with a minimally coupled scalar field:
\begin{eqnarray}
S = \int d^{4}x \sqrt{-g} \left[ \frac{1}{16\pi}R - \frac{1}{2} (\nabla \phi)^{2} - V(\phi) \right].
\end{eqnarray}
We consider a quadratic potential,
\begin{equation}
V(\phi) = V_{0} + \frac{1}{2} m^{2} \phi^{2}
\end{equation}
and we only consider the small field limit: $|\phi| \ll 1$. Then one can approximate that the Hubble parameter $H\simeq\sqrt{8\pi V_{0}/3}$ is a constant.

The basic motivation of the stochastic approach comes from this observation: one can define a minimal length $\lambda_{c} \gtrsim H^{-1}$ or a maximal momentum $q_{c} \propto 1/\lambda_{c}$, and if the potential is broad and the field value is sufficiently small, then we can treat the wavelengths longer than $\lambda_{c}$ as a classical and locally homogeneous field, while we can treat wavelengths shorter than $\lambda_{c}$ as a Gaussian random noise to the longer wavelengths. In other words, our universe is homogeneous up to the length scale $\lambda_{c}$. Therefore, in this approximation, we can use the homogeneous metric ansatz: $ds^{2} = - dt^{2} + a^{2}(t)\textbf{dx}^{2}$ to describe shorter region than $\lambda_{c}$. Now, the homogeneous universe (up to the length $\lambda_{c}$) will behave by a homogeneous quantum fluctuation of the scalar field; the field now only depends on time and looks like a random walking, or looks like a stochastic process.

In the following subsections, we will sketch the systematics of the stochastic approach.

\subsection{Quantum fields in de Sitter space}

By introducing the conformal time $d\eta = dt/a$ and defining a scalar field as $\varphi \equiv a \phi$,
one has the equation of motion for $\varphi$ as
\begin{equation}
\varphi '' - \nabla^2 \varphi + \left( a^2 m^2 - \frac{a''}{a}   \right) \varphi = 0,
\end{equation}
where a prime is the derivative with respect to $\eta$.
This has the general solution
\begin{equation}
\varphi(\eta, \textbf{x}) = \int \frac{d^3 \textbf{k}}{(2 \pi)^{3/2}}
\left[ a_{\textbf{k}} \varphi_k(\eta) + a^{\dagger}_{-\textbf{k}} \varphi^*_k(\eta) \right] e^{i \textbf{k} \cdot \textbf{x}}
\end{equation}
where $\varphi_k$ satisfies
\begin{equation}\label{eq:varphi_t}
\varphi''_k + \left( k^2 + a^2 m^2 -\frac{a''}{a}  \right) \varphi_k = 0
\end{equation}
and is normalized such that
\begin{equation}
\varphi_k{\varphi^*}'_k - \varphi'_k\varphi^*_k = i.
\end{equation}
The quantization condition is reduced to
\begin{equation}
\left[a_{\textbf{k}}, a^{\dagger}_{\textbf{l}}    \right] = \delta^3(\textbf{k} - \textbf{l}).
\end{equation}
Note that the $k$ is a comoving momentum; the physical momentum can be represented by $q = k /a$.

In de Sitter space, $H$ is a constant. Hence, $a = e^{Ht}$ and
\begin{equation}
\eta = - \frac{1}{a H}.
\end{equation}
Then Equation~(\ref{eq:varphi_t}) becomes
\begin{equation}\label{eq:varphi_eta}
\varphi''_k +  k^2 \varphi_k + \frac{1}{\eta^{2}} \left( \frac{m^2}{H^2} - 2  \right) \varphi_k = 0.
\end{equation}
On the scales well inside the horizon, $-k \eta \rightarrow \infty$, this equation reduces to
\begin{equation}
\varphi''_k +  k^2 \varphi_k = 0,
\end{equation}
which has the normalized solution
\begin{equation}
\varphi_{k} = \frac{1}{\sqrt{2k}} \left( A_{k} e^{-ik\eta} + B_{k} e^{ik\eta} \right), \;\;\;\; \left|A_{k} \right|^{2}-\left|B_{k} \right|^{2} = 1.
\end{equation}
If the inflationary expansion has been going on for a sufficiently long time, the scalar field should be in the usual flat space vacuum state on scales well inside the horizon. Therefore, we should take $B_{k}=0$ so that $a_{k}$ and $a_{k}^{\dag}$ correspond to the usual flat space annihilation and creation operators, and the state should be $|0\rangle$ where $a_{k}|0\rangle=0$. We are free to take $A_{k}=1$ to get
\begin{equation}
\varphi_k \rightarrow \frac{1}{\sqrt{2k}} e^{-ik \eta}
\end{equation}
as $-k\eta \rightarrow \infty$.

Since the solution of Equation~(\ref{eq:varphi_eta}) should be reduced into the above in the sub-horizon limit,
the solution is expressed as
\begin{equation}
\varphi_k = e^{i \left( \nu + \frac{1}{2} \right)\frac{\pi}{2} }
\sqrt{\frac{\pi}{4k}} \sqrt{-k \eta} H^{(1)}_{\nu}(-k \eta),
\end{equation}
where
\begin{equation}\label{eq:nu}
\nu = \sqrt{\frac{9}{4}-\frac{m^2}{H^2}}.
\end{equation}
Hence, on the scales well outside the horizon, $-k \eta \rightarrow 0$
the asymptotic form of the solution is given as
\begin{equation}
\varphi_k \rightarrow e^{i \left( \nu -\frac{1}{2} \right)\frac{\pi}{2}}
\left[ \frac{2^{\nu}\Gamma(\nu)}{2^{\frac{3}{2}}\Gamma\left(\frac{3}{2} \right)}  \right] \frac{1}{\sqrt{2k}} (-k \eta)^{\frac{1}{2}-\nu}.
\end{equation}

This allows us to rewrite the super-horizon Fourier modes, i.e., those with $k \ll aH$, as
\begin{equation}
a_{\textbf{k}} \varphi_{k}(\eta)+a^{\dag}_{-\textbf{k}} \varphi^{*}_{k}(\eta) = b_{\textbf{k}} \left[ \frac{2^{\nu}\Gamma(\nu)}{2^{\frac{3}{2}}\Gamma\left(\frac{3}{2} \right)}  \right] \frac{1}{\sqrt{2k}} (-k \eta)^{\frac{1}{2}-\nu},
\end{equation}
where
\begin{equation}
b_{\textbf{k}}= e^{i\left(\nu-\frac{1}{2} \right)\frac{\pi}{2}} a_{\textbf{k}}+e^{-i\left(\nu-\frac{1}{2} \right)\frac{\pi}{2}} a^{\dag}_{-\textbf{k}}.
\end{equation}
Therefore, we can observe
\begin{equation}
\left[b_{\textbf{k}}, b^{\dagger}_{\textbf{l}}    \right] = 0,
\end{equation}
and so the super-horizon Fourier modes are classical Gaussian random variables with
\begin{equation}
\langle0|b_{\textbf{k}}b^{\dagger}_{\textbf{l}} |0\rangle = \delta^{3} (\textbf{k}-\textbf{l}).
\end{equation}

To summarize, if we consider only the super-horizon Fourier modes ($k \ll a H$, or $q \ll H$; equivalently, $\lambda \gg H^{-1}$), we can describe the universe as a \textit{locally} homogeneous universe and the scalar field stochastically behaves as the classical Gaussian random walking. This is the reason why we call this method the stochastic approach.

\subsection{Dispersion of fields with a quadratic potential}

Now we calculate the dispersion of $\phi$:
\begin{equation}
\langle 0 | \phi^2  |0 \rangle = \frac{1}{a^2} \langle 0 | \varphi^2  |0 \rangle
= \frac{1}{a^2} \int \frac{d^3k}{(2\pi)^3} |\varphi_k |^2.
\end{equation}
This integral in itself has a divergence as $k \rightarrow \infty$. Therefore, we need to introduce some sort of ultraviolet cutoff, $k_c$. This is consistent with our stochastic approach; we consider only the super-horizon Fourier modes so that $k < k_{c} \sim a H$. Then, the momentum cutoff $k_{c}$ or physical momentum cutoff $k_{c}/a$ corresponds to the smoothing length scale (the minimal length of the relevant wave) $\lambda_{c} \propto 1/q_{c}$ that determines how to coarse-grain the momentum space. Hence, the integration can be expanded as follows:
\begin{equation}\label{eq:phi2primitive}
\langle 0 | \phi^2 |0 \rangle =  \frac{1}{a^2} \int_0^{k_c} \frac{d^3k}{(2\pi)^3} |\varphi_k |^2 = \left( \frac{H}{2\pi}\right)^2 \left(\frac{2^{\nu}\Gamma(\nu)}{2^{\frac{3}{2}}\Gamma\left( \frac{3}{2}\right)} \right)^2
\frac{1}{3-2 \nu} \left(\frac{k_{c}}{aH} \right)^{3-2\nu}.
\end{equation}
Then, by assuming $m^2/H^2 \ll 1$, this is expanded as
\begin{equation}
\langle 0 | \phi^2 |0 \rangle \simeq \frac{3 H^{4}}{8 \pi^{2} m^{2}} \left[ 1 + \frac{2 m^{2}}{3 H^{2}} \ln \left( \frac{k_{c}}{\zeta}\frac{1}{aH} \right)+ \mathcal{O}\left( \frac{m}{H} \right)^{3} \right],
\end{equation}
where
\begin{equation}
\ln \zeta = \ln 2 + \frac{1}{9\pi} + \frac{2}{3\pi} \frac{\Gamma'\left(\frac{3}{2}\right)}{\Gamma\left(\frac{3}{2}\right)},
\end{equation}
and hence $\zeta \simeq 2.45$. This can be absorbed by a new momentum scale $\tilde{k}_{c} = k_{c}/\zeta$. For future convenience, we define $\mu^{2} = m^{2}/V_{0}$ and rewrite it as follows:
\begin{eqnarray}\label{eq:stocha}
\langle \phi^2 \rangle \simeq \frac{8}{3 \mu^{2}}V_{0} \left[ 1 + \frac{\mu^{2}}{4 \pi^{2}} \ln \frac{\tilde{k}_{c}}{aH} + \mathcal{O}\left( \mu^{3} \right) \right].
\end{eqnarray}

Although the dispersion $\langle \phi^{2} \rangle$ is a specific quantity, this is indeed a representative value of the stochastic approach, since the probability distribution of the statistical ensemble can be approximated, in general,
\begin{eqnarray}
P(\phi,t) \simeq \frac{1}{\sqrt{2\pi \langle (\phi(t)-\phi_{\mathrm{cl}}(t))^{2} \rangle}} \exp \left[ -\frac{(\phi(t)-\phi_{\mathrm{cl}}(t))^{2}}{2\langle (\phi(t)-\phi_{\mathrm{cl}}(t))^{2} \rangle} \right],
\end{eqnarray}
where $\phi_{\mathrm{cl}}$ is the field value of the purely classical trajectory, if $\left|\phi - \phi_{\mathrm{cl}}\right| \ll 1$. Therefore, in the stationary state, the dispersion $\langle (\phi-\phi_{\mathrm{cl}})^{2} \rangle$ (around the local minimum $\phi_{\mathrm{cl}}=0$, $\langle (\phi-\phi_{\mathrm{cl}})^{2} \rangle = \langle \phi^{2} \rangle$) will be a time-independent value, and, for a small field value $\phi$, the dispersion will determine the approximate distribution of the fields.

\section{\label{sec:euc}Euclidean approach}

\subsection{Hawking-Moss instantons}

The wave function of the universe to describe the ground state of the universe \cite{Hartle:1983ai} is
\begin{eqnarray}
\Psi[h_{\mu\nu}, \chi] = \int_{\partial g = h, \partial \phi = \chi} \mathcal{D}g\mathcal{D}\phi \;\; e^{-S_{\mathrm{E}}[g,\phi]},
\end{eqnarray}
where $h_{\mu\nu}$ and $\chi$ are the boundary values of the Euclidean metric $g_{\mu\nu}$ and the matter field $\phi$ which are the integration variables, and the integration is over all non-singular geometries with a single boundary. Here, we consider the Euclidean action
\begin{eqnarray}
S_{\mathrm{E}} = -\int d^{4}x \sqrt{+g} \left( \frac{1}{16\pi}R - \frac{1}{2} (\nabla \phi)^{2} - V(\phi) \right).
\end{eqnarray}

In the minisuperspace approximation
\begin{eqnarray}
ds_{\mathrm{E}}^{2} = d\tau^{2} + \rho^{2}\left(\tau\right) d\Omega_{3}^{2},
\end{eqnarray}
the Euclidean action becomes
\begin{eqnarray}
S_{\mathrm{E}} = 2 \pi^{2} \int d\tau \left[ -\frac{3}{8\pi} \left( \rho \dot{\rho}^{2} + \rho \right) + \frac{1}{2}\rho^{3} \dot{\phi}^{2} + \rho^{3} V(\phi) \right].
\end{eqnarray}

We use the steepest-descent approximation, and we only consider the on-shell solutions to count the probability from the path-integral. We solve the classical equations of motion for Euclidean and Lorentzian time directions:
\begin{eqnarray}
\ddot{\phi} &=& - 3 \frac{\dot{\rho}}{\rho} \dot{\phi} \pm V',\\
\ddot{\rho} &=& - \frac{8 \pi}{3} \rho \left( \dot{\phi}^{2} \pm V \right),
\end{eqnarray}
where the upper sign is for the Euclidean time and the lower sign is for the Lorentzian time.
The on-shell Euclidean action is
\begin{eqnarray}
S_{\,\mathrm{E}} = 4\pi^{2} \int d \tau \left( \rho^{3} V - \frac{3}{8 \pi} \rho \right).
\end{eqnarray}

When $V'(\phi_{0})=0$, we can find an analytic solution:
\begin{eqnarray}
\phi &=& \phi_{0}\\
\rho &=& \frac{1}{H} \sin H \tau,
\end{eqnarray}
where $H=\sqrt{8\pi V(\phi_{0})/3}$. This satisfies the regular initial conditions:
\begin{eqnarray}\label{eq:initial}
\rho(0) = 0, \;\;\; \dot{\rho}(0) = 1, \;\;\; \dot{\phi}(0) = 0.
\end{eqnarray}
We want to analytically continue to the Lorentzian manifold using $\tau = X + i t$. Then at the turning point $\tau = X$, we have to impose
\begin{align} \label{eq:paste}
\rho(t=0) = \rho(\tau=X), \;\;\;\; \dot{\rho}(t=0)=i\dot{\rho}(\tau=X),\\
\label{eq:paste2}
\phi(t=0) = \phi(\tau=X), \;\;\;\; \dot{\phi}(t=0)=i\dot{\phi}(\tau=X),
\end{align}
from the analyticity of complex functions. Unless $\dot{\rho}(\tau=X) = \dot{\phi}(\tau=X) = 0$, we should have complex valued functions for $\phi$ and $\rho$ for the Lorentzian time $t$. Therefore, we can analytically continue at $\tau = \pi/2H$; then we can maintain real valued functions for all $\tau$ and $t$.

In this case, the on-shell action becomes
\begin{eqnarray}
S_{\mathrm{E}} = - \frac{3}{16 V(\phi_{0})}.
\end{eqnarray}
If this instanton mediates a tunneling from a local minimum $\phi_{\mathrm{m}}$ to a local maximum $\phi_{\mathrm{M}}$ of a potential, then we can define the tunneling probability:
\begin{eqnarray}
P \cong \exp \left( \frac{3}{8 V(\phi_{\mathrm{M}})} -  \frac{3}{8 V(\phi_{\mathrm{m}})} \right).
\end{eqnarray}
This is known as the Hawking-Moss instanton \cite{Hawking:1981fz}.

\subsection{Fuzzy instantons}

We can generalize these real-valued instantons to complex valued functions. Of course, for a long Lorentzian time, we expect that all functions should become real. This condition is called the \textit{classicality condition} \cite{Hartle:2007gi}. Because of the analytic continuation to complex functions, the action is, in general, complex, so that
\begin{equation}
\Psi[a,\chi] = A[a,\chi] e^{i S[a,\chi]}
\label{eq:class}
\end{equation}
with $A,S$ real.
If the rate of change of $S$ is much greater than that of $A$,
\begin{equation} \label{eqn:classicality}
|\nabla_I A(q)|\ll |\nabla_I S(q)|, \qquad I=1,\ldots n,
\end{equation}
where $I$ denotes the canonical variables, then the wave function describes almost classical behavior.

We require initial conditions (Equation~\ref{eq:initial}) for regularity and, at the junction time $\tau = X$, we paste $\rho(\tau)$ and $\phi(\tau)$ to $\rho(t)$ and $\phi(t)$ as in Equations~(\ref{eq:paste}) and (\ref{eq:paste2}). The remaining initial conditions are the initial field value $\phi(0) = \phi_{0} e^{i\theta}$, where $\phi_{0}$ is a positive value and $\theta$ is a phase angle. After fixing $\phi_{0}$, by tuning the two parameters $\theta$ and the turning point $X$, we will try to satisfy the classicality condition \cite{Hartle:2007gi}\cite{Hwang:2011mp}. If there exists a classical history, then we can calculate a meaningful probability for a classical universe.

\subsection{Fuzzy instantons with a quadratic potential}

One can study the probability via fuzzy instantons by an approximate way \cite{Lyons:1992ua}: around the local minimum of the potential
\begin{equation}
V(\phi) = \frac{1}{2} m^{2} \phi^{2};
\end{equation}
the starting point is the following approximate solutions of the equations of motion:
\begin{eqnarray}
\phi \simeq \phi_{0} + i \frac{m}{3} \sqrt{\frac{3}{4\pi}}\tau, \;\;\; \rho \simeq \sqrt{\frac{3}{4\pi}}\frac{i}{m\phi_{0}} \exp \left(-i\sqrt{\frac{4\pi}{3}}m\phi_{0}\tau + \frac{1}{6}m^{2}\tau^{2} \right),
\end{eqnarray}
in which the scalar field $\phi$ slowly rolls.
If the scalar field rolls more slowly, then we can further approximate
\begin{eqnarray}
\label{approx2}\phi \simeq \phi_{0}, \;\;\; \rho \simeq \sqrt{\frac{3}{4\pi}}\frac{1}{m\phi_{0}} \sin \left(\sqrt{\frac{4\pi}{3}}m\phi_{0}\tau \right).
\end{eqnarray}

We choose the integration contour in two steps. $(1)$ We integrate in the Euclidean time direction from $\tau = 0$ to $\tau = \sqrt{3\pi}/4m \phi_{0} \equiv X$ so that the imaginary part of $\phi$ vanishes. $(2)$ At the turning point $\tau = X$, we turn to the Lorentzian time direction.

Using this contour of integration, the Euclidean action can be calculated. Note that, if the classicality condition is valid, the real part of the action $S_{\mathrm{E}}$ picks up the biggest contribution during the Euclidean time integration. Using Equation~(\ref{approx2}), we can calculate the Euclidean action and the result is
\begin{eqnarray}\label{eq:fuzzyaction}
S^{(1)}_{\mathrm{E}} = 4\pi^{2} \int_{0}^{X} \left( \rho^{3} V - \frac{3}{8 \pi} \rho \right) d \tau \simeq - \frac{3}{8 m^{2} (\phi_{0})^{2}} \sim - \frac{3}{16 V(\phi_{0})}.
\end{eqnarray}
Therefore, as the vacuum energy becomes smaller and smaller, the probability gets larger and larger. This qualitative result is confirmed in more detailed calculations by Hartle, Hawking and Hertog \cite{Hartle:2007gi}.

\begin{figure}
\begin{center}
\includegraphics[scale=0.65]{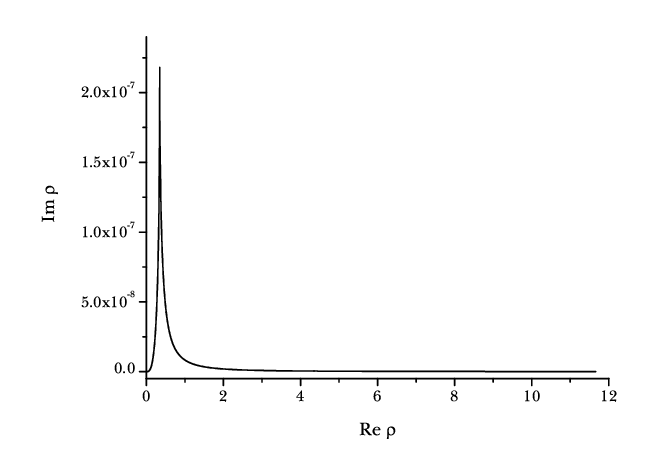}
\includegraphics[scale=0.65]{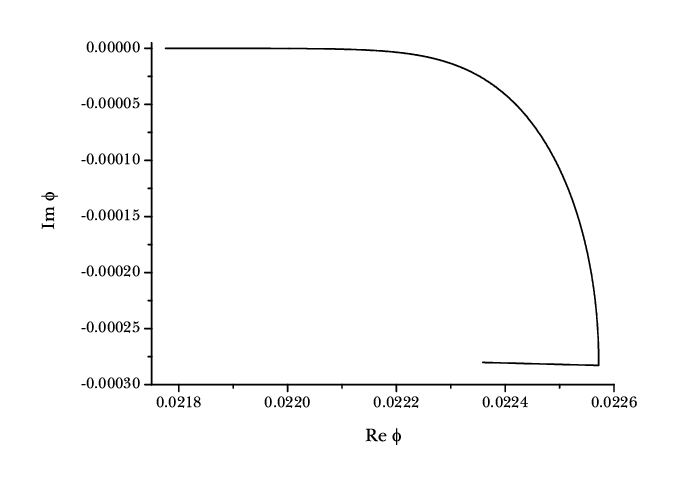}
\caption{\label{fig:rho}An example of a fuzzy instanton solution with $\mu^{2}=0.2$ and $\mu\phi_{0}=0.02$. The imaginary part decreases to zero and eventually classicalized. Here, the cusp is the turning point.}
\end{center}
\end{figure}
\begin{figure}
\begin{center}
\includegraphics[scale=0.65]{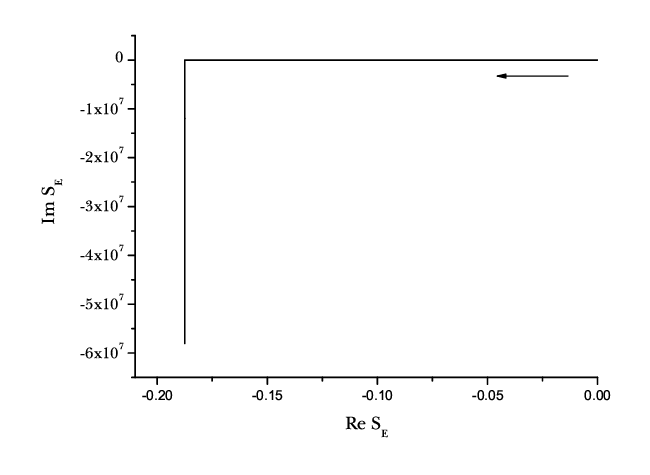}
\includegraphics[scale=0.65]{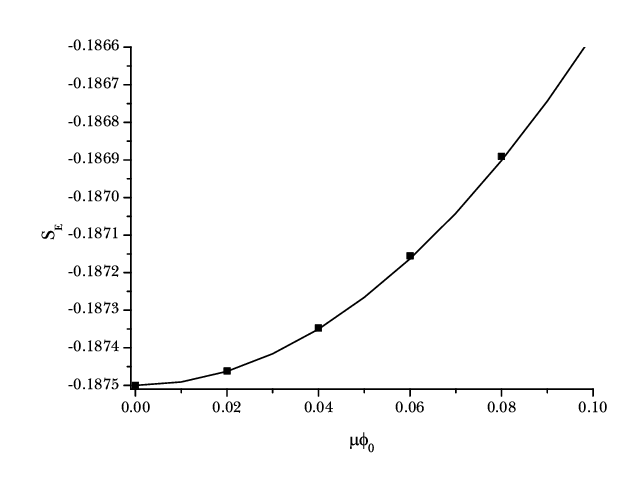}
\caption{\label{fig:action}Left: The real part and the imaginary part of the Euclidean action for $\mu^{2}=0.2$ and $\mu\phi_{0}=0.02$. One can see that the solution is eventually classicalized. Here, the cusp is the turning point. Right: We can write the Euclidean action for classicalized histories as a function of $\mu\phi_{0}$. The black curve is the approximate formula (Equation~\ref{eq:fuzzyaction}) for the slow-roll limit.}
\end{center}
\end{figure}

We can confirm this expectation by using a numerical searching (the same methods used in \cite{Hwang:2011mp}, see Appendix for details). Figure~\ref{fig:rho} shows an example of a classicalized history. Both the imaginary parts of $\rho$ and $\phi$ eventually decrease to zero. Left of Figure~\ref{fig:action} is the real part and the imaginary part of the Euclidean action. This shows that after the turning point, the imaginary part of the Euclidean action quickly varies, while the real part of the Euclidean action that contributes to the probability extremely slowly varies; this is a clear signature of the classicality. After we find classicalized histories for various $\phi_{0}$, we can see the probability distribution as a function of $\mu\phi_{0}$ (Right of Figure~\ref{fig:action}). This fits well with the analytic calculations.

\section{\label{sec:the}Conceptual descriptions}

In previous sections, we discussed two independent approaches: the stochastic approach and the Euclidean approach. In both cases, we assumed Einstein gravity with a minimally coupled scalar field and a quadratic potential. What we compare in detail is the dispersion of the field $\langle \phi^2 \rangle$ in both of the approaches. However, now the crucial question is this: \textit{Why do we have to compare two results, while two approaches seem to describe different physical situations by different methods?} Does the comparison make sense?

There is no trivial way to justify this, because we do not know the totality of the stochastic universe (multiverse) and also we do not know the full quantum gravity (even for the Euclidean version). However, it is not unreasonable to believe (or, assume as a working hypotheses) that the Euclidean path integral describes a ground state wave function of the universe, while the stochastic approach describes the probability distribution of the scalar field due to the stochastic random walking, or the statistical distribution of various ensembles that is generated by stochastic inflation. As in simple examples of Brownian motion, the stationary state of the probability distribution due to the stochastic random walking can correspond to the ground state of the Schrodinger equation \cite{LeB}.

This is the basic motivation of the comparison. Let us discuss it in further detail.

\subsection{Expectations}

We state again on our general intuitions between two approaches:
\begin{enumerate}
\item The \textit{stochastic approach} describes the probability distribution of quantum states, or statistical distribution of many ensembles. After a sufficiently long time, the distribution will approach the \textit{equilibrium}/\textit{stationary state}\footnote{We regard that the stationary state is the equilibrium or $t \rightarrow \infty$ limit, although there can be some time-independent state that is not the $t \rightarrow \infty$ limit.}. In this limit, we think that the system approaches the \textit{ground state}.
\item The \textit{Euclidean path integral} describes the \textit{ground state wave function}. The ground state wave function is a superposition of many \textit{histories}. Each history is approximated by \textit{instantons}.
\end{enumerate}

If we know the sound quantum descriptions, then we \textit{hypothesize} the two expectations:
\begin{description}
\item[Expectation 1:] If there is a state for certain physical properties in the stationary state, then \textit{there is an instanton} for the physical properties that has the same probability.
\item[Expectation 2:] The modulus square of the ground state wave function corresponds to the probability distribution of the stationary state.
\end{description}

\subsection{Ambiguities}

For a realistic comparison, we need to clarify more details on each side. The physical meaning of $\langle \phi^2 \rangle$ is much clearer in the stochastic side; it is a statistical distribution of the scalar field around the local minimum. However, what about the Euclidean side?

\begin{figure}
\begin{center}
\includegraphics[scale=0.45]{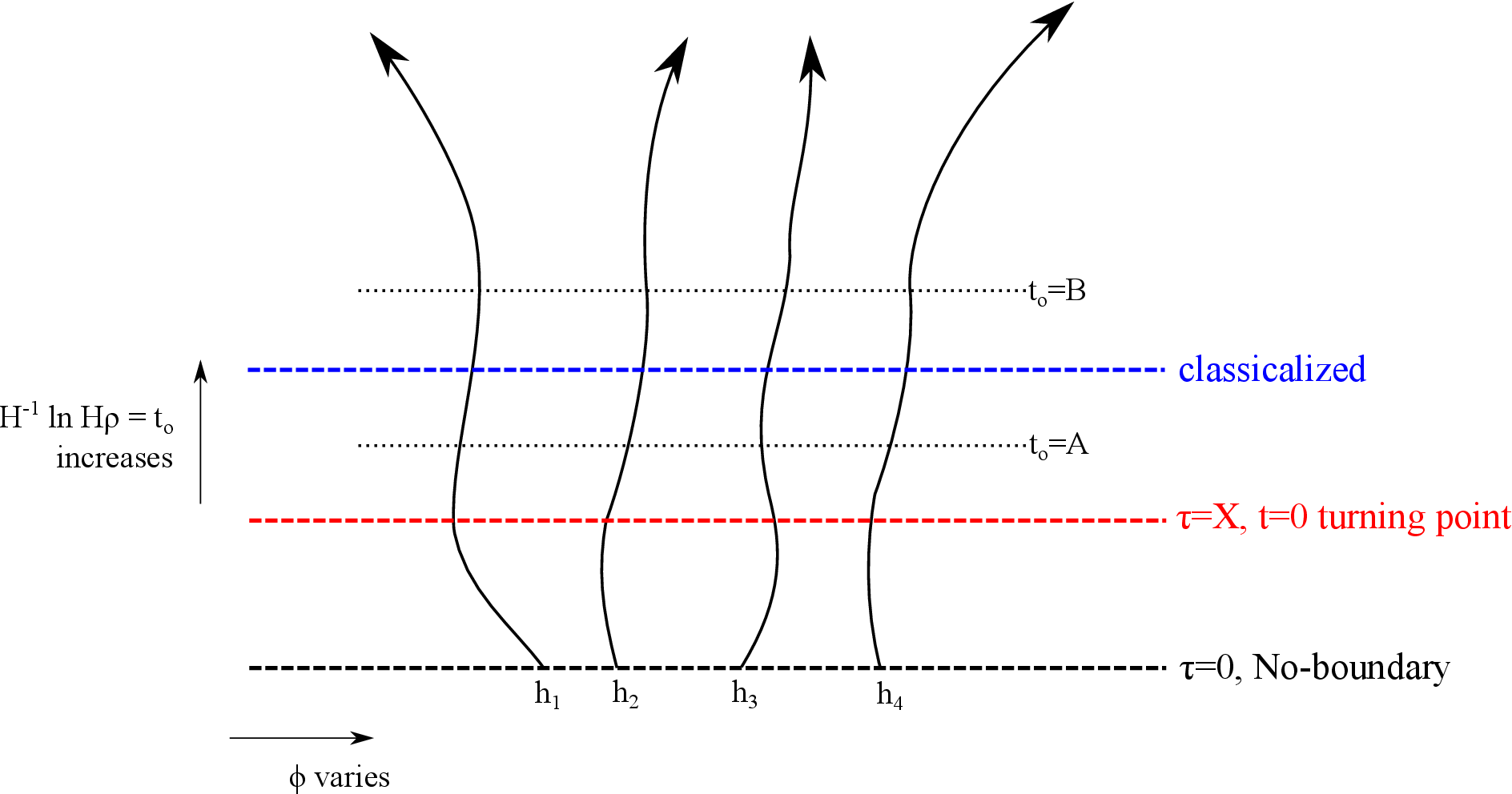}
\caption{\label{fig:scheme}A schematic diagram of $\Psi[\rho,\phi]$.}
\end{center}
\end{figure}

Figure~\ref{fig:scheme} represents the schematic diagram that happens in the wave function. The wave function is approximated by superposition of classical histories (instantons): thick black arrows $h_{1}$, $h_{2}$, ..., etc. Each instanton begins from no-boundary $\tau=0$ (black dashed line), turns to the Lorentzian time $\tau=X, t=0$ (red dashed line), and is eventually classicalized (around the blue dashed line). Vertical and horizontal lines conceptually represent the field space (minisuperspace). For de Sitter-like spaces, we can approximate the minisuperspace as vertically increasing as the scale factor or $t_{o} = H^{-1} \ln H \rho$ increases while horizontally increasing as $\phi$ varies.

The probability is determined for each history. We can write the probability as a function of $\Psi[\rho, \phi]$ at a certain slice $t_{o}=A$ or $t_{o}=B$, i.e., $\Psi[\phi]_{t_{o}}$. Although the probability is fixed for each history, the functional dependence between $\Psi[\phi]_{t_{o}=A}$ and $\Psi[\phi]_{t_{o}=B}$ can be different. This is easy to see, if the field slowly rolls; there is a one-to-one correspondence between the time and the scalar field or the scale factor, and it can be simplified as follows:
\begin{eqnarray}
\Psi[\phi,\rho]=\Psi[\phi[\rho]]=\Psi[\phi]_{t_{o}}.
\end{eqnarray}

Therefore, what we will see for a given quadratic potential is the following. In the stochastic approach, there are many ensembles. Each observer will experience Brownian random walking; however, in the thermal equilibrium, the number of events for a certain field value $\phi$ can be maintained at an almost constant value. In this case, $\langle \phi^{2} \rangle$ is the dispersion of the statistical distribution. On the other hand, in the Euclidean approach, there are many histories. Each history corresponds a creation of a universe that has a certain field value $\phi(t)$ from nothing. For all histories, we can cut a certain slice. The slice should be labeled by a canonical variable, e.g., $t_{o}\simeq H^{-1} \ln H \rho$; we call $t_{o}$ the observing time. Then, we can write the wave function as a function of canonical variables. In this case, $\langle \phi^{2} \rangle$ is the dispersion of the modulus square of the wave function.

Here, one interesting observation is that in both of approaches, there is ambiguity for $\langle \phi^{2} \rangle$. In the stochastic approach, there is the smoothing scale $k_{c}$ dependence. On the other hand, in the Euclidean approach, there is the observing time $t_{o}$ dependence.

In the present paper, we will compare the dispersion $\langle \phi^{2} \rangle$ in both approaches, stochastic and Euclidean approaches, not only to the leading order, but also the second order and more.
\begin{enumerate}
\item From Expectation 1, we can find the nature of the corresponding instanton in the Euclidean side, if two approaches give the same results.
\item From Expectation 2, in addition, if the results of both approaches are the same, then this will be reduced to the comparison between the ambiguities of $k_{c}$ and $t_{o}$. We can see whether it physically makes sense or not; if it physically makes sense, then we can be sure that we are going in the right direction.
\end{enumerate}

\section{\label{sec:com}Comparison between stochastic and Euclidean approaches}

\subsection{\label{sec:1st}The leading order}

Note that one can rewrite the probability as the Gaussian form (this is valid for a small field region):
\begin{eqnarray}
\ln P(\phi) \simeq - \frac{\phi^{2}}{2 \langle \phi^{2} \rangle}.
\end{eqnarray}
We can compare with the Euclidean action (Equation~(\ref{eq:fuzzyaction})):
\begin{eqnarray}
\ln P(\phi) \simeq -2 S_{\mathrm{E}} \simeq \frac{3}{8V_{0}}\frac{1}{1+ \mu^{2} \phi^{2} /2} \simeq \frac{3}{8V_{0}} \left( 1 - \frac{1}{2}\mu^{2} \phi^{2} \right).
\end{eqnarray}
Note that the constant term is the normalization constant in the exponential. Therefore, we can identify that
\begin{eqnarray}
\langle 0| \phi^{2} |0 \rangle \simeq \frac{8}{3 \mu^{2}}V_{0} = \frac{3H^{4}}{8\pi^{2} m^{2}}.
\end{eqnarray}
Thus, this corresponds to the leading term of the result of the stochastic approach (Equation~(\ref{eq:stocha})).

It is interesting to focus on the extreme case $m = 0$. Then, the dispersion diverges, and the probability distribution as a solution of the Fokker-Planck equation spreads over the field space. This is intuitively true, since there should be no special position. On the Euclidean side, if $m = 0$, then the probability for any field value is the same with $S_{\mathrm{E}}=-3/16V_{0}$. Hence, if we consider the normalization, again, the probability distribution spreads over the field space. Therefore, we can conclude that two approaches are consistent even for the trivial case $m = 0$.

Note that the result in Equation~(\ref{eq:fuzzyaction}) is similar to the probability for the Hawking-Moss instantons. In addition, people already know that the probability of the Hawking-Moss instanton is indeed a stationary solution of the Fokker-Planck equation that comes from the stochastic Langevin equations. Therefore, the leading term correspondence is not a new observation.

However, up to now, people did not worry about this aspect of the correspondence between two pictures (the stochastic approach and the Euclidean approach). If the ground state wave function corresponds to the probability distribution of stochastic inflation (the solution of the Fokker-Planck equation), then for a given state of the solution of the Fokker-Planck equation, there should be a corresponding \textit{instanton} in the Euclidean side, since the Euclidean wave function is approximated by the superposition of instantons. Now, our question is this: \textit{What is the corresponding instanton for a quadratic potential?}

One candidate is the real valued instantons (Hawking-Moss instantons). However, the real valued instanton is impossible unless there is a point $\tau_{0}$ such that $\dot{\rho}(\tau_{0})=\dot{\phi}(\tau_{0})=0$. It is possible only at a local minimum\footnote{Some exceptional cases happen when the slow-roll condition is not satisfied \cite{Lee:2012qv}.}. Therefore, the real valued instantons are not suitable.

The answer is that \textit{it should be a complexified instanton} (or so-called fuzzy instanton). If the stochastic approach is more physical than the Euclidean approach (the Euclidean approach is a rather mathematical one), this shows that \textit{there is a corresponding reality for a fuzzy instanton}. Up to now, it was unclear whether the fuzzy instanton has the physical meaning or it is just a mathematical illusion. Now, the leading term correspondence shows that the fuzzy instanton is not an illusion but a physical reality.

We can extend our assertions, not only to the leading term (that is already known), but also to the next term (that is not yet investigated) to search a non-trivial correspondence.

\subsection{\label{sec:2nd}The second order}

For a given quadratic potential ($V_{0}+m^{2}\phi^{2}/2$, $\mu^{2}=m^{2}/V_{0}$), one can define a rescaling:
\begin{eqnarray}
d\tau &\rightarrow& \frac{d\tau}{\sqrt{V_{0}}},\\
\rho &\rightarrow& \frac{\rho}{\sqrt{V_{0}}}.
\end{eqnarray}
Then we can obtain the action rescaling
\begin{eqnarray}
S_{\mathrm{E}} \rightarrow \frac{S_{\mathrm{E}}}{V_{0}}
\end{eqnarray}
and the rescaled potential
\begin{eqnarray}
V(\phi) \rightarrow \frac{V(\phi)}{V_{0}}.
\end{eqnarray}
Therefore, without loss of generality, we can choose $V_{0} = 1$, and finally one can restore the results. In this subsection, we will choose $V_{0}=1$.

\subsubsection{Euclidean approach: Numerical observations}

\begin{figure}
\begin{center}
\includegraphics[scale=0.7]{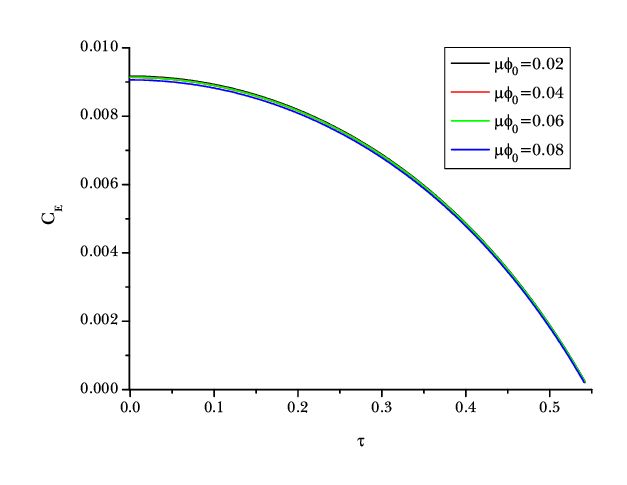}
\includegraphics[scale=0.7]{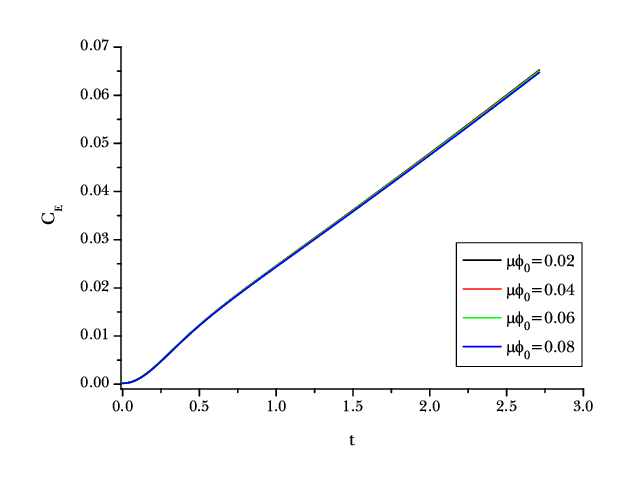}
\caption{\label{fig:xdependence}$C_{\mathrm{E}}$ as a function of observing time $\tau$ and $t$, by fixing $\mu^{2}=0.2$ and varying $\mu \phi_{0} = 0.02, 0.04, 0.06, 0.08.$ The gradient of $C_{\mathrm{E}}(t)$ is approximately $0.02387.$}
\end{center}
\end{figure}

\begin{figure}
\begin{center}
\includegraphics[scale=0.7]{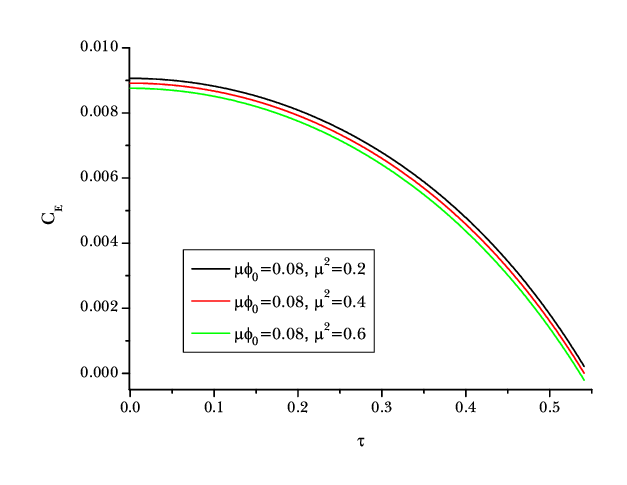}
\includegraphics[scale=0.7]{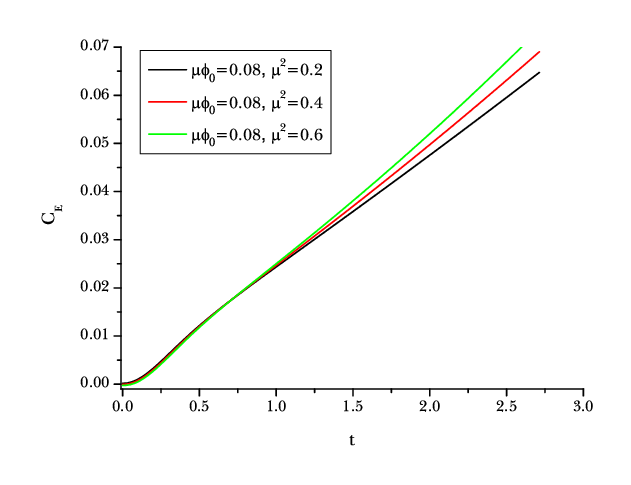}
\caption{\label{fig:mudependence}$C_{\mathrm{E}}$ as a function of observing time $\tau$ and $t$, by fixing $\mu \phi_{0}=0.8$ and varying $\mu^{2} = 0.2, 0.4, 0.6.$ The gradient of $C_{\mathrm{E}}(t)$ is approximately $0.02387, 0.02496$, and $0.02621$, respectively.}
\end{center}
\end{figure}

To obtain the second-order contribution, we suggest that the action looks like
\begin{eqnarray}\label{eq:actionform}
S_{\mathrm{E}} = -\frac{3}{16 \left( 1 + \mu^{2}\phi^{2}(t_{o})/2 \right)} +C_{\mathrm{E}}(t_{o}) \mu^{2}  \left( \mu\phi(t_{o}) \right)^{2}.
\end{eqnarray}
From the numerical observations, we can show that $C_{\mathrm{E}}(t_{o})$ only depends on $t_{o}$ and does not depend on $\mu$ or $\phi_{0}$ (Figures~\ref{fig:xdependence} and \ref{fig:mudependence}), at least, for $\mu \ll 1$ and $\mu \phi \ll 1$.
In addition, numerical observations show that $C_{\mathrm{E}}$ almost \textit{linearly} depends on the observing time $t_{o}$, where $t_{o}$ is the \textit{observation time after the turning point}. The intercept of this linear approximation is nearly zero.

$S_{\mathrm{E}}$ does not depend on the choice of $t_{o}$, and, hence, when $\phi(t_{o})$ varies, $C_{\mathrm{E}}(t_{o})$ should also vary. From the Euclidean action, we estimate that
\begin{eqnarray}
\langle \phi^{2} \rangle \left.\right|_{t_{o}} \simeq \frac{8}{3\mu^{2}} \left( 1-\frac{32}{3}C_{\mathrm{E}}(t_{o}) \mu^{2} \right).
\end{eqnarray}
According to numerical observations, the dependence on $t_{o}$ is approximately linear after the turning point and the numerical fitting is as follows:
\begin{eqnarray}
C_{\mathrm{E}} \simeq 0.02 \times t_{o}.
\end{eqnarray}
This approximation is still fine if $t_{o} \gtrsim 1/H \sim X \sim 0.5$.


This is not an accident. Differentiating both sides of Equation~(\ref{eq:actionform}) by $t_{o}$,
\begin{eqnarray}
\dot{C}_{\mathrm{E}} &\simeq& -\frac{3 \dot{\phi}}{16 \mu^{2} \phi} \left( 1 + \frac{32}{3}C_{\mathrm{E}}\mu^{2} \right)\\
&\simeq& \frac{1}{16H} \left( 1 + \frac{32}{3} C_{\mathrm{E}} \mu^{2} \right).
\end{eqnarray}
Here, we assumed the slow-roll condition $\dot{\phi} \simeq -V'/3H$. Note that $1/16H = 3H/128\pi$. Therefore, perturbatively one can solve
\begin{eqnarray}
C_{\mathrm{E}} \simeq \frac{3}{128 \pi} \sqrt{\frac{8\pi}{3}} (t - C) \simeq 0.0216 \times (t - C),
\end{eqnarray}
where $C$ is a constant. By comparing the numerical calculations, we can choose $t-C = t_{o}$.

\subsubsection{Euclidean approach: Analytic discussion}

From the numerical observations, the Euclidean action is approximately $S_{\mathrm{E}}=-3/16V(\phi_{\mathrm{T}})$, where $\phi_{\mathrm{T}}$ is the field observed at the turning point. Using this, one can discuss in a formal way. Note that we can use the equations of motion in the slow-roll limit:
\begin{eqnarray}
H^{2} &\simeq& \frac{8\pi}{3}V,\\
\dot{\phi} &\simeq& - \frac{V'}{3H}.
\end{eqnarray}

At the observing time $t_{o}$, the potential varies:
\begin{eqnarray}
V(\phi) \simeq V(\phi_{\mathrm{T}}) + V' \dot{\phi} \times t_{o},
\end{eqnarray}
and hence, the Euclidean action can be expanded by
\begin{eqnarray}
S_{\mathrm{E}} &\simeq& -\frac{3}{16 V(\phi_{\mathrm{T}})}\\
&\simeq& -\frac{3}{16 V(\phi)} \left( 1 + \frac{V'}{V} \dot{\phi} \times t_{o} \right)\\
&\simeq& -\frac{3}{16 V(\phi)} + \frac{t_{o}}{16 H} \mu^{4}\phi^{2}.
\end{eqnarray}

Therefore, compare to Equation~(\ref{eq:actionform}), we obtain $C_{\mathrm{E}} = t_{o}/16H$, and it is the same result of the previous section. Therefore, the universality of $C_{\mathrm{E}}$ comes from the fact that we use the small $\mu^{2}$ limit; the field slowly varies. However, the interpretation that $t_{o}$ is the time from the turning point came from numerical observations.

\subsubsection{Stochastic approach}

From the stochastic approach, we can write as follows:
\begin{eqnarray}
\langle \phi^{2} \rangle \simeq \frac{8}{3\mu^{2}} \left( 1-\frac{32}{3}C_{\mathrm{S}} \mu^{2} \right),
\end{eqnarray}
where
\begin{eqnarray}
C_{\mathrm{S}} \equiv -\frac{3}{128 \pi} \ln \frac{\tilde{k}_{c}}{a H}.
\end{eqnarray}
We can represent $\tilde{k}_{c}/H=\exp H(t-\Delta t)$, where $\Delta t$ is a certain constant that we can choose freely. Then
\begin{eqnarray}
\ln \frac{\tilde{k}_{c}}{a H} \simeq - H \Delta t = - \sqrt{\frac{8\pi}{3}} \Delta t,
\end{eqnarray}
and, hence,
\begin{eqnarray}
C_{\mathrm{S}} \simeq \frac{3}{128\pi} \sqrt{\frac{8\pi}{3}} \Delta t \simeq 0.0216 \times \Delta t.
\end{eqnarray}

\subsubsection{Interpretations}

Our claim is that
\begin{eqnarray}
C_{\mathrm{S}} = C_{\mathrm{E}}.
\end{eqnarray}
In other words,
\begin{eqnarray}
\Delta t = t_{o}.
\end{eqnarray}
The left hand side purely comes from the analysis of the quantum field theory in de Sitter space. The right hand side comes from the Euclidean quantum gravity, and it corresponds to the time after the turning point of a certain universe.

\begin{figure}
\begin{center}
\includegraphics[scale=0.45]{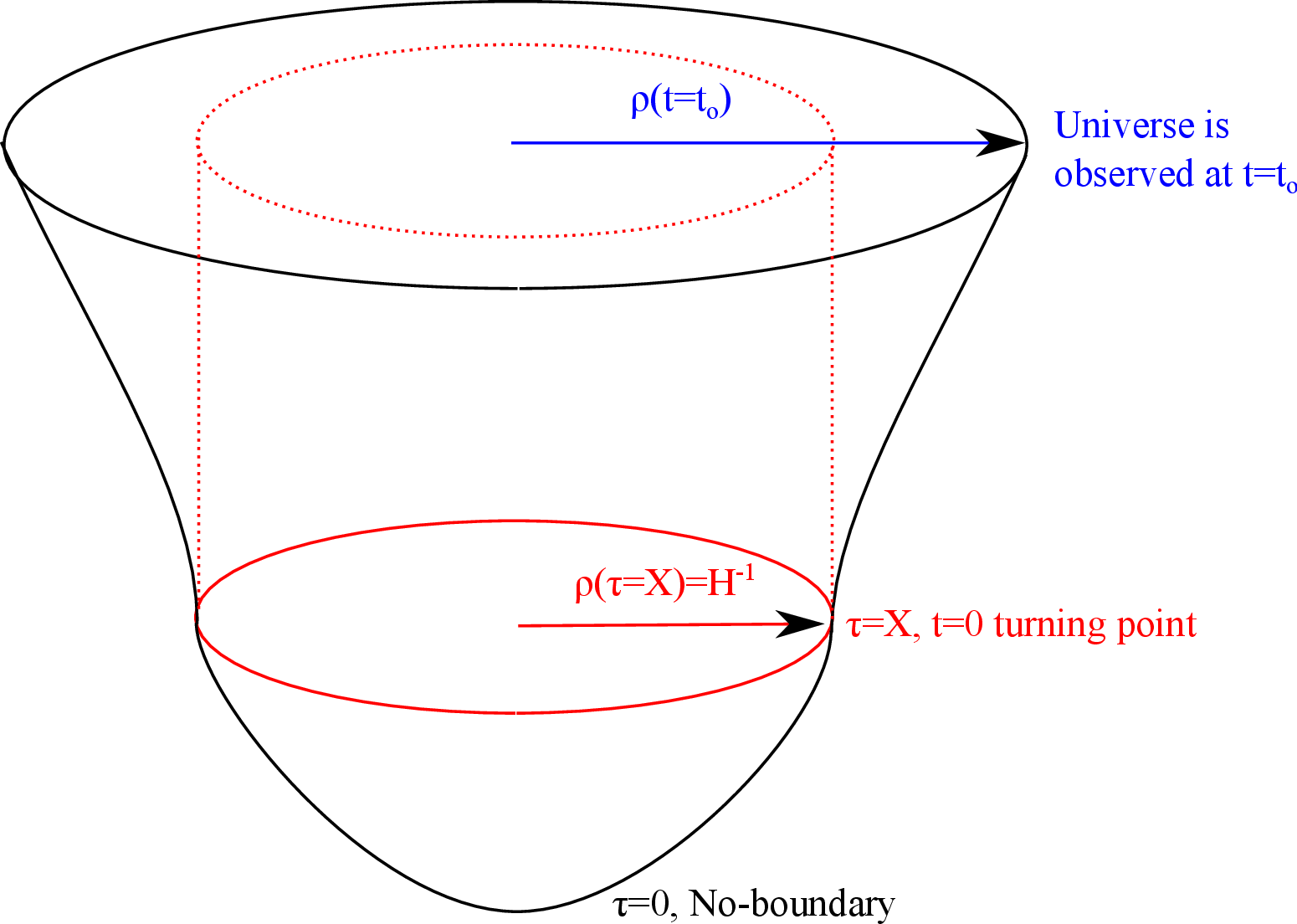}
\caption{\label{fig:interpretation}Physical interpretation of the cutoff scale.}
\end{center}
\end{figure}

Now, we give physical interpretations (Figure~\ref{fig:interpretation}). In terms of scale factors, the smoothing scale is
\begin{eqnarray}
\frac{aH}{\tilde{k}_{c}} \simeq \frac{\rho(t_{o})}{\rho(\tau=X)},
\end{eqnarray}
or
\begin{eqnarray}
\frac{1}{\tilde{q}_{c}} \simeq \rho(t_{o}),
\end{eqnarray}
where $\tilde{q}_{c}\equiv\tilde{k}_{c}/a$ is the physical cutoff scale of the momentum space. In the stochastic approach, if a certain quantum fluctuation has a shorter wave length than $\lambda_{c} \propto 1/\tilde{q}_{c}$, then we can ignore it; the universe is homogeneous up to the length scale $\lambda_{c}$. On the other hand, in the Euclidean approach, the instanton is homogeneous up to the size of the universe $\rho(t_{o})$. Now what we can say is that
\begin{eqnarray}
\lambda_{c} \sim \rho({t_{o}}).
\end{eqnarray}

This is a new observation for Euclidean quantum gravity. The Euclidean instanton comes from the no-boundary state, and the instanton is compact in itself. After the Lorentzian analytic continuation, the volume of the entire universe is still finite, unless the time is infinite. What is the meaning of the finite size of the universe $\rho$? What is the outside of $\rho$? After the comparison to the stochastic approach, we can say clearer statement: $\rho$ corresponds our \textit{coarse-graining lattice} when we approximate that the entire universe is locally (not globally) homogeneous.

\subsection{\label{sec:3rd}Large $\mu^{2}$ limit}

Finally, we consider the situation $\mu^{2} > 6\pi$. Then, $\nu$ (Equation~\ref{eq:nu}) becomes imaginary. In the stochastic side, if $\nu$ is imaginary, it can be interpreted that longer wavelength modes begin to oscillate and the long wavelength modes no longer dominate, so results depend strongly on any smoothing scale. Therefore, once $\nu$ turns imaginary, the long wavelength modes can no longer be considered approximately classical.

On the other hand, according to Hartle, Hawking, and Hertog \cite{Hartle:2007gi}, one may find an interesting correspondence in the Euclidean side.
Let us first observe the perturbative discussion in \cite{Hartle:2007gi}. To study the first perturbative level, we fix the background metric as follows:
\begin{eqnarray}
\rho(\tau) = \frac{1}{H} \sin H\tau &~& \;\;\;\;\;\;\;\; \mathrm{for~} 0 \leq \tau \leq X,\\
\rho(t) = \frac{1}{H} \cosh H t &~& \;\;\;\;\;\;\;\; \mathrm{for~} 0 \leq t.
\end{eqnarray}
Then, the equation for the scalar field becomes
\begin{eqnarray}
\ddot{\phi} + 3 H \cot H\tau \dot{\phi} - m^{2} \phi &=& 0 \;\;\;\;\;\;\;\; \mathrm{for~} 0 \leq \tau \leq X,\\
\ddot{\phi} + 3 H \coth H t \dot{\phi} + m^{2} \phi &=& 0 \;\;\;\;\;\;\;\; \mathrm{for~} 0 \leq t.
\end{eqnarray}
In general, the solution $G(\tau)$ with the property $G(0)=1$ and $\dot{G}(0)=0$ becomes $G(\tau) = F(a,b,c,z(\tau))$, where $F$ is the hypergeometric function and
\begin{eqnarray}
a &=& \frac{3}{2} \left( 1 + \sqrt{1- \frac{\mu^{2}}{6\pi}} \right),\\
b &=& \frac{3}{2} \left( 1 - \sqrt{1- \frac{\mu^{2}}{6\pi}} \right),\\
c &=& 2,\\
z(\tau) &=& \frac{1 - \cos H\tau}{2}.
\end{eqnarray}
Then, using the function $G(\tau)$, one can construct the function $\phi$ with a certain final field value $\phi(\tau_{\mathrm{fin}})=\chi$ or $\phi(t_{\mathrm{fin}})=\chi$ by
\begin{eqnarray}
\phi(\tau) &=& \chi \frac{G(\tau)}{G(\tau_{\mathrm{fin}})},\\
\phi(t) &=& \chi \frac{G\left(\pi/2H + i t \right)}{G\left(\pi/2H + i t_{\mathrm{fin}}\right)}.
\end{eqnarray}
Therefore, whenever we impose $\chi$ as a real valued function, the initial phase angle
\begin{eqnarray}
\theta = - \arctan \frac{\mathfrak{Im}~G(\pi/2H + i t_{\mathrm{fin}})}{\mathfrak{Re}~G(\pi/2H + i t_{\mathrm{fin}})}
\end{eqnarray}
is determined. In other words, if the condition
\begin{eqnarray}\label{eq:condition}
\frac{\mathfrak{Im}~G(\pi/2H + i t)}{\mathfrak{Re}~G(\pi/2H + i t)} \rightarrow \mathrm{constant}
\end{eqnarray}
is satisfied for a sufficiently large $t$, then we can redefine the phase angle $\theta$ so that we can construct a realized field solution. However, if it does not converge to a constant, then there is no hope to tune $\theta$ to construct a realized field solution.

Note that the equations for the real part $\mathfrak{Re}~G$ and the imaginary part $\mathfrak{Im}~G$ along the large Lorentzian time effectively become
\begin{eqnarray}\label{eq:G1}
\mathfrak{Re}~\ddot{G} + 3 H\; \mathfrak{Re}~\dot{G} + m^{2}\; \mathfrak{Re}~G &=& 0,\\
\label{eq:G2}
\mathfrak{Im}~\ddot{G} + 3 H\; \mathfrak{Im}~\dot{G} + m^{2}\; \mathfrak{Im}~G &=& 0.
\end{eqnarray}
To satisfy the reality condition, Equation~(\ref{eq:condition}), we require that Equations~(\ref{eq:G1}) and (\ref{eq:G2}) should not be oscillatory; while if these are oscillatory, Equation~(\ref{eq:condition}) cannot be satisfied. Note that Equations~(\ref{eq:G1}) and (\ref{eq:G2}) are just simple damped harmonic oscillators, and these are overdamped if $\mu^{2} < 6\pi$, underdamped if $\mu^{2} > 6\pi$, and critically damped if $\mu^{2} = 6\pi$. The underdamped condition corresponds to the case when the real and imaginary parts are both oscillatory and difficult to classicalize.

Therefore, if $\mu^{2} > 6\pi$, then there is a cutoff $\phi_{c}$ such that the \textit{classicality condition} is unstable around the local minimum. Now, we summarize the logical flow:
\begin{enumerate}
\item For the underdamped region $\mu^{2} > 6 \pi$, the real part and the imaginary part of the field oscillate around the local minimum. Therefore, there is no classical history, or the classicality is inevitably related to the oscillation of the fields.
\item For the same region $\mu^{2} > 6 \pi$, longer wavelength modes cannot be regarded classical and the local (Hubble scale) \textit{homogeneity} is violated.
\end{enumerate}

Therefore, for highly curved potentials $\mu^{2} > 6 \pi$, the classicalization is related to the oscillation of the fields. The oscillation of fields will be related to \textit{inhomogeneity} and \textit{instability of classicality}. We leave further details for future work.

\section{\label{sec:con}Conclusion}

In this paper, we compare the two approaches. One is the quantum field theory in a de Sitter background; the so-called stochastic inflation. The other is the Euclidean quantum gravity using fuzzy instantons. The former describes a probability distribution of a random walking field and a solution of the Fokker-Planck equation, while the latter is the wave function of the ground state of the Wheeler-DeWitt equation. We expect that the stationary state of the probability distribution should correspond to the modulus square of the ground state wave function.

For practical comparison, we compared the dispersion of the field, where this is indeed representative in both approaches. We integrate three important conclusions:
\begin{description}
\item[Leading order:] They match well (Section~\ref{sec:1st}). This implies that the fuzzy (complex-valued) instanton is indeed the corresponding instanton on the Euclidean side, while there is no real-valued instanton.
\item[Second order:] Still two approaches are the same (Section~\ref{sec:2nd}). This implies that the size of the universe on the Euclidean approach corresponds to the smoothing scale size of the stochastic approach. In the stochastic approach, the universe is homogeneous up to the length scale $\lambda_{\mathrm{c}}$, and this is the size of the scale factor of the Euclidean approach $\rho$.
\item[Large $\mu^{2}$ limit:] If the curvature of the potential around the local minimum is sufficiently large, then the potential should satisfy the underdamped condition. Then, the classicality in the Euclidean approach is related to the oscillation of the field, and this is related to inhomogeneity in the stochastic approach (Section~\ref{sec:3rd}).
\end{description}

One interesting comparison is that the stochastic inflation can realize all the possible quantum states during inflation. Therefore, this can realize statistical ensembles. The Euclidean side introduces a wave function, while the stochastic side can correspond to a statistical system. If our knowledge about quantum mechanics, the so-called \textit{Born rule}\footnote{We use Born's rule as follows: the modulus square of a wave function corresponds the probability distribution of statistical ensembles.}, is sound, then the present work corresponds to Born's rule of the \textit{quantum gravitational version}.

Our comparison shows that there are non-trivial correspondences between the Euclidean approach and the stochastic approach. In this paper, we go over the previous discussions in the two points. First, fuzzy instantons are not only mathematical, but also corresponds to something in nature. In other words, fuzzy instantons are realized by stochastic inflation. Second, the finite size of the universe of the instantons in the no-boundary measure indeed has physical meaning from the stochastic inflation. This is the boundary of the homogeneous approximation, a grid size of the stochastic approach. These can be good indirect evidences that the Euclidean quantum gravity and quantum cosmology go in the right direction. In addition, we expect that there should be more relations even beyond the slow-roll limits. We leave the detailed analysis for future work.

\section*{Acknowledgment}

The authors would like to thank Hanno Sahlmann, Robert Brandenberger, Erick Weinberg, Soo A Kim, and Wonwoo Lee for helpful discussions. DY, BHL, and DH were supported by the National Research Foundation grant (2005-0049409) funded by the Korean government through the Center for Quantum Spacetime(CQUeST) of Sogang University. EDS and HZ were supported by the National Research Foundation grant (2009-0077503) funded by the Korean government.

\section*{Appendix: Numerical searching algorithm}

There is no real-valued instanton except the local minimum of the quadratic potential. However, we can find a complex-valued instanton for general field values. Once we generalize to complex-valued functions, then we have eight initial conditions: the real part and imaginary part of $\rho(0)$, $\phi(0)$, $\dot{\rho}(0)$, and $\dot{\phi}(0)$. Among these conditions, we already fix six of them, since we require the regularity of $\tau=0$: $\rho(0)=0$, $\dot{\rho}=1$, and $\dot{\phi}=0$. Then, there remain two initial conditions: $\phi(0)=\phi_{0} e^{i\theta}$, where $\phi_{0}$ is the modulus of the initial field position and $\theta$ is the phase angle. In addition, we have to choose a turning point $X$ from the Euclidean time $\tau$ to the Lorentzian time $t$. Therefore, for a complex-valued instanton with a given $\phi_{0}$, we still have undefined two-dimensional degrees of freedom: $(\theta, X)$.

We have to impose the classicality for a given complex-valued instanton, since it should return to a classical universe for a sufficiently large $t$. This is implemented by tuning $\theta$ and $X$. To find a classical fuzzy instanton, therefore, we need parameter tuning for $\theta$ and $X$; the purpose of the tuning is to minimize
\begin{equation}
\frac{|\nabla_I A(q)|}{|\nabla_I S(q)|}, \qquad I=1,\ldots n.
\end{equation}
For this purpose, we first try to find candidates of $\theta$ and $X$ that minimize an objective function (this can be chosen freely for our technical conveniences). For example, we minimize the objective function,
\begin{eqnarray}
F_{\phi_{0}} \left[\theta, X \right] \equiv \int_{T_{1}}^{T_{2}} \left| \frac{\mathcal{L}^{\mathfrak{Re}}_{\phi_{0}}[\theta, X](t)}{ \mathcal{L}_{\phi_{0}}[\theta, X](t)}\right| dt.
\end{eqnarray}
Here, $\mathcal{L}_{\phi_{0}}[\theta, X](t)$ is the Lagrangian with a given initial field modulus $\phi_{0}$. $T_{1}$ and $T_{2}$ are sufficiently large time values when we measure the Lagrangian and are introduced for technical conveniences. Of course, in general, the candidates $\theta$ and $X$ do not imply the classicality. Hence, we have to check again whether $|\phi^{\mathfrak{Im}}/\phi^{\mathfrak{Re}}| \ll 1$ and $|\rho^{\mathfrak{Im}}/\rho^{\mathfrak{Re}}| \ll 1$ for a sufficiently large $t$.

To summarize, the searching algorithm finds the best candidate of the fuzzy instanton. We should check whether this candidate is indeed a classicalized fuzzy instanton or not. If it is so, then we can say that there is a fuzzy instanton, it has a certain action, etc. If one of the conditions $|\phi^{\mathfrak{Im}}/\phi^{\mathfrak{Re}}| \ll 1$ or $|\rho^{\mathfrak{Im}}/\rho^{\mathfrak{Re}}| \ll 1$ does not hold, then we should conclude that there is no fuzzy instanton for a given initial condition $\phi_{0}$.

To optimize the objective function $F_{\phi_{0}}$, we used the well-known genetic algorithm. The technical details are discussed in \cite{Hwang:2011mp}. After we find all the fuzzy instantons, we can discuss the action $S_{\mathrm{E}}$ as a function of $\phi_{0}$. After we fix a fitting form such as Equation~(\ref{eq:actionform}), we can discuss further details on the dispersion.

\end{document}